\documentstyle[12pt,epsfig,amstext]{article}

\begin{document}

\title{ Tsallis Entropy and the transition to scaling in fragmentation } 

\author{ Oscar Sotolongo-Costa$^{1,2}$, Arezky H. Rodriguez$^1$ and G. J. Rodgers$^2$}

\maketitle
\begin{center}
\footnotesize{1.- Department of Theoretical Physics, Havana University, Habana 10400, Cuba.\\
 2.- Department of Mathematical Sciences, Brunel University,\\
Uxbridge, Middlesex UB8 3PH, UK.\\}
\end{center}
\begin{abstract}
By using the maximum entropy principle with Tsallis entropy 
we obtain a fragment size distribution function which undergoes a transition to scaling.
This distribution function
reduces to those obtained by other authors using Shannon entropy.
The treatment is easily generalisable to any process of fractioning with
 suitable constraints.  
\end{abstract}

\noindent 05.40.Fb, 24.60.-k

\newpage

As a result of developments in materials science, combustion technology, geology and many 
other fields of research, there has been an increase of interest
in the problem of fragmentation of objects. Within this general 
field there
 is a collection
of papers  \cite {oscar,matsu,madeira,rmf} where a transition occurs
from a ``classical'' distribution of fragments ({\em e.g.} log-normal or
 Rossin-Ramler- like)
to a power law distribution. This transition has not been
 adequately explained in terms of any general principles, although in \cite {oscar}
 the representation of the fragmentation process in terms of  percolation on a Bethe lattice
leads to a transition to a power law in the distribution of fragment sizes.\\
Some attempts have been made to derive the fragment size distribution function 
from the maximum entropy principle \cite{li,ruso}, subject 
to some constraints which mainly came from physical considerations about the 
fragmentation phenomena. The resulting fragment distribution 
function describes the distribution of sizes of the fragments in a regime in which 
scaling is not present.\\
As scaling invariably occurs  when the energy of the
 fragmentation process is high,  this suggests that the analysis is only applicable to
low energies.
However, the maximum entropy principle is completely universal and has 
an almost unlimited range 
of application. Consequently we would expect to be able to use it to describe
the transition to scaling as the energy of the fracture grows.\\

The expression for the Boltzmann-Gibbs entropy $S$ ({\em{e.g.}} Shannon's form) is given by

\begin{equation}
S=-k \sum_{i=1}^{W} p_ilnp_i, \label{eq:1}
\end{equation}

where $p_i$ is the probability of finding the system in the microscopic state i, 
 $k$ is Boltzmann's constant, and $W$ is the total number of microstates.\\
This has been 
shown to be restricted to the domain of validity of Boltzmann-Gibbs (BG) statistics. 
These statistics seem to describe nature when the effective microscopic interactions and
 the microscopic memory are short ranged \cite{rbf}.
The process of violent fractioning, like that of droplet
 microexplosions in combustion chambers, blasting and shock fragmentation 
with high energies, etc, leads to the existence of long-range correlations between {\em{all}}
parts of the object being fragmented.\\ 
Fractioning is a paradigm of non extensivity, since the fractioning
object can be considered as a collection of parts which, after division, have 
an entropy larger than that of their union {\em{i.e}} if 
we denote by $A_i$ the parts or fragments in which the object has  been divided,
 its entropy $S$ obeys $S(\bigcup A_i) < \sum_i S(A_i)$, defining 
a ``superextensivity'' \cite{rbf} in this system . This suggests that it may be
 necessary to use non-extensive
statistics, instead of the BG statistics. This kind of theory has already been
proposed by Tsallis \cite{tsallis}, who postulated a generalized form of entropy,
given by

\begin{equation}
S_q=k \frac{1-\int_0^\infty p^q(x)dx }{q-1}.  \label{eq:2}
\end{equation}
The integral 
runs over all admissible values of the magnitude $x$ and $p(x)dx$  is the 
probability of the system being in a state between $x$ and $x+dx$.
This entropy can also be expressed as\\
 $$S_q= \int p^q(x)ln_q p(x)dx .$$
  The generalized logarithm
$ln_q(p)$ is defined in \cite{rbf} as
\begin{equation}
ln_q(p)= \frac{p^{1-q} -1}{1-q},  
\end{equation}
where q is a real number. It is straighforward to see that $S_q\rightarrow S$ when $q\rightarrow 1$, 
recovering BG statistics.\\
In this paper we use the entropy in eq.\ref{eq:2} to consider the problem of atomization of 
liquid fuels.

The atomized drops in a low pressure regime follow a Nukiyama-Tanasawa-like
 distribution of sizes, a 
particular case of the Rossin-Ramler distribution \cite{williams}. As we already pointed out
in \cite{madeira}, this distribution tends to a power-law, revealing scaling as the
atomization pressure grows. Incidentally, this is essentially  the same behavior as that
 reported in experiments on 
 falling glass rods \cite{matsu},  mercury drops, \cite{oscar} and on blasting 
oil drops \cite{rmf}.
So, let us maximise $\frac{S_q}{k}$ given by eq.\ref{eq:2}. If we denote 
the volume of a drop by $V$ and  some typical volume characteristic of the distribution by $V_m$,
we can define a dimensionless volume $v=\frac{V}{V_m} $. Then, the constraints to impose are

\begin{equation}
\int_0 ^\infty p(v)dv =1 ,  \label{eq:3}
\end{equation}
{\em{i.e.}}, the normalization condition. 
The other condition to be imposed is mass conservation. But as the 
system is finite, this condition will lead to a very sharp decay in the asymptotic behavior 
of the droplet size distribution function (DSDF) for large sizes of the fragments.
 Consequently, we will impose a more general condition, 
like the ``q-conservation'' of the mass, in the form: 

\begin{equation}
\int_0^\infty vp^q(v)dv=1,   \label{eq:4}
\end{equation}

which reduces to the ``classical'' mass conservation when $q=1$.

Equations \ref{eq:3} and \ref{eq:4} are the constraints to impose in order to derive the 
DSDF using the method of Lagrange multipliers 
by means of the construction of the function: 

\begin{equation}
L(p_i;\alpha_1;\alpha_2)= S_q- \alpha_1 \int_0^\infty p(v)dv
+\alpha_2 \int_0^\infty p^q(v)vdv \label{eq:6}
\end{equation}
The Lagrange multipliers  $\alpha_1$ and $\alpha_2$  are determined from
eqs.\ref{eq:3} and \ref{eq:4}.
The extremization of $L(p_i; \alpha_1; \alpha_2)$ leads to:
\begin{equation}
p(v)dv= C(1+(q-1)\alpha_2v)^{-\frac{1}{q-1}}dv \label{eq:7}
\end{equation}
where the constant $C$ is given by $$C=\frac{q-1}{q} \alpha_1 ^{\frac{1}{q-1}}.$$\\
This is a DSDF expressed as a function of  the volume of the droplets. 
It is convenient to formulate
 the problem in terms of a DSDF
as a function of the dimensionless radius of the droplets $r=v^{1/3}$. Then the 
probability density is:

\begin{equation}
f_q(r)=3Cr^2[1+(q-1)\alpha_2r^3)]^{-\frac{1}{q-1}}. \label{eq:8}
\end{equation}

To obtain the DSDF the range 
of admissible values of $q$ is $1<q<2$. This range of values of 
$q$ also guarantees the adequate power law behavior of eq.\ref{eq:8}, since its asymptotic 
behavior
for large $r$ is

\begin{equation}
f_q(r)\sim \frac{1}{r^{\alpha +1}}, \label{eq:13}
\end{equation}

where $\alpha$ is the generic power law exponent, $\alpha=3\frac{2-q}{q-1}$.
Also, if $q\rightarrow 1$, eq.\ref{eq:8} leads to:

\begin{equation}
f(r)=3r^2exp(-r^3), \label{eq:14}
\end{equation}

which is the a Nukiyama-Tanasawa DSDF, a particular case of the Rossin- Ramler distribution.
This distribution has been  previously obtained in \cite{li}.\\
 Then, the DSDF that we have obtained reproduces the actual behavior
of fragments in the process of breaking. It is easy to realise that the 
above viewpoint is applicable not only to atomization, but to any process of fractioning.\\
For a given regime of breakage, generally identified as that of the lowest energy of breakage,
the fragment distribution function can be deduced through the maximum entropy principle
using BG statistics. This low energy regime of breakage is such that the correlations between 
the different
 parts of the object are short-ranged.
As the energy of breakage increases, long-range correlations
 become more and more important, which
makes it necessary to introduce the Tsallis entropy as a generalization of the 
Shannon entropy.\\
 Thus,
we have confirmed that BG statistics cannot be applied to all fragmentation regimes 
and Tsallis entropy can be used to describe the transition into scaling. In this respect, the 
parameter q, which determines the ``degree of nonextensivity''
 of the statistics, can be related to an effective temperature of breakage.
As far as we know, this is the first formulation in terms of general principles
that leads to a DSDF which exhibits a transition to scaling.\\
This work was partially supported by the ``Alma Mater'' contest, Havana University. 
One of us (O.S) is grateful to the Department of Mathematical Sciences
 of Brunel University for
 kind hospitality and the Royal Society, London for financial support.

\end{document}